\documentclass[twocolumn,showpacs,preprintnumbers,amsmath,amssymb,aps]{revtex4}

\usepackage{graphicx}

\newcommand{\lb}{{<}}
\newcommand{\rb}{{>}}
\newcommand{\ft}{{\cal F}}

\begin{document}

\title{Simulations of Spinodal Nucleation in Systems with 
Elastic Interactions}

\author{C.\ J.\ Gagne}
\affiliation{Department of Physics, Clark University,
Worcester, MA 01610}

\author{W.\ Klein}

\altaffiliation[Permanent Address: ] {Department of Physics and Center
for Computational Science, Boston University, Boston, MA 02215}

\affiliation{Theoretical Division, Los Alamos National Laboratory, Los Alamos,
NM 87545}

\author{T.\ Lookman}
\affiliation{Theoretical Division, Los Alamos National Laboratory, Los Alamos,
 NM 87545}

\author{A.\ Saxena}
\affiliation{Theoretical Division, Los Alamos National Laboratory, Los Alamos,
 NM 87545}

\author{H.\ Gould}
\affiliation{{}Department of Physics, Clark University,
Worcester, MA 01610}

\begin{abstract}
Systems with long-range interactions quenched into a metastable state 
near the pseudospinodal exhibit nucleation that is qualitatively 
different than the classical nucleation observed near the coexistence 
curve.  We have observed
nucleation droplets in our Langevin simulations of a two-dimensional
model of martensitic transformations and have determined that
the structure of the nucleating droplet differs from the stable martensite 
structure. Our results, together with 
experimental measurements of the phonon dispersion curve, allow us to 
predict the nature of the droplet. These results have implications for 
nucleation in many solid-solid transitions and the structure of the final 
state.
\end{abstract}
\bigskip
\noindent 

\date{\today}
\maketitle

An outstanding challenge in the physics of solid-solid phase transformations
is to understand the nucleation and growth of the transformed phase in strain 
based
materials such as martensites, ferroelectrics, and multiferroics.
Although ideas based on classical nucleation theory have been invoked to
describe nucleation phenomena in these materials~\cite{borg,
nishiyama,oshima,olson,reid,vzg}, it is possible
that the presence of long-range elastic forces
substantially influences the probability of nucleation and
subsequent growth that determines processing and material behavior.
For example, the transformation susceptibility determined by
nucleation is one of the basic factors for the hardenability of steels.
Even though heterogeneous nucleation, which is sensitive to the distribution
of  appropriate structures in the parent phase from which product phase
nuclei may be  triggered, has long been considered important for
martensites~\cite{cohen}, homogeneous nucleation also has  been observed
if the
transformational driving force is sufficiently large. However, due to the
difficulty of experimentally determining 
nucleation droplet structure in martensites~\cite{borg},
very little is quantitatively known 
about the morphology of the nucleating droplet or ``embryo,'' including its
size  distribution or nucleation rate for homogeneous or heterogeneous
nucleation.  The aim of this work is to probe for the first time
the nature of the nucleating droplet and associated fluctuations 
by performing mesoscale simulations using realistic nonlinear models for 
martensites. We find that the classical theory does not accurately describe 
the structure of the nucleating droplet, but that concepts associated with 
nucleation near a spinodal~\cite{uk, kleinley} account for the observed
droplet morphology.  In addition, our work provides the basis for
investigating nucleation in systems where  long-range elastic forces
crucially determine the morphology.

Long-range elastic interactions are important in strain based materials, 
such as martensites, and result from the requirement of compatibility
of strain components which is necessary to preserve the continuity of the 
elastic media~\cite{slsb,kksw}. We refer to systems with long-range interactions as 
near-mean-field, and use several characteristics of mean-field theory to 
study nucleation in these systems.  Mean-field systems have a well-defined 
spinodal, the limit of metastability; near-mean-field systems have a 
pseudospinodal, which becomes better defined as the range of interaction 
increases~\cite{heermann}.

Spinodal nucleation, that is, nucleation close to the pseudospinodal,
is predicted to produce a ramified droplet with a small
amplitude~\cite{uk, kleinley}. 
The droplet need not have the same structure as the stable 
phase~\cite{klsh,kleinpre}, unlike in classical nucleation where the droplet 
is compact and has the same structure as the stable
phase~\cite{lang}. Spinodal 
nucleation has been observed in molecular dynamic simulations of simple
models~\cite{yang,cbssk}, but the concepts have never been tested on more
realistic representations of materials such as martensites. 

We model a martensitic transformation using a 
Ginzburg-Landau (GL) free energy of the
form~\cite{slsb, kksw,klsh}
\begin{subequations}
\label{e:GL}
\begin{align}
F[\phi]& = 
F_0 + F_{\rm{grad}}
 + F_{\rm{cs}} \label{1a} \\
F_0& = \!\int_{0}^{L}\!\! d^2r \big[ 
\tau \phi^2 - 2 \phi^4 + \phi^6 \big] \label{1b}\\ 
F_{\rm{grad}}& = \!\int_{0}^{L}\!\! d^2r 
\left[ \frac{a}{4} (\nabla\phi)^2 + \frac{b}{8} (\nabla^2\phi)^2 
\right] \label{1c} \\ 
F_{\rm{cs}}& = \!\int_{0}^{L}\!\! d^2r 
\left[ \frac{A_1}{2} e_1^2 + \frac{A_2}{2} e_2^2 \right] \label{1d} \\
& \approx \!\int_{0}^{L}\!\! d^2r d^2r' U(\vec r - \vec r')
e^{-|\vec r - \vec r'|/R} \phi(\vec r) \phi(\vec r') , \label{1e}
\end{align}
\end{subequations}
where $\phi$ is the deviatoric strain, $e_1$ is the shear strain, $e_2$ is the 
compressional strain, $U(\vec \rho = \vec r - \vec r')$ is the
Fourier transform of
\begin{equation}
\hat U(\vec k) = \frac{A_1}{2} \frac{(k_x^2-k_y^2)^2}{\left[
k^4 + 8 \frac{A_1}{A_2} k_x^2k_y^2 \right]},
\label{e:ubulk}
\end{equation}
$\tau= (\theta - \theta_c)/(\theta_0 - \theta_c)$, $\theta$ is the
dimensionless temperature,
$\theta_c$ is the critical temperature where the $\phi=0$ austenite minimum 
of $F_0$ disappears, $\theta_0$ is the temperature where the three minima of
$F_0$ are degenerate, and $R$ is the range of the interaction. For metals $R$ is quite 
large. The quantity $F_{\rm cs}$ can be written as nonlocal surface and 
bulk terms in $\phi(\vec r)$ by using the St.\ Venant compatibility 
equations~\cite{slsb}. If the width of the interface of the droplet scales as 
the correlation length $\xi$, the surface term can be neglected 
for $\xi>> R$~\cite{klsh,long}. The exponential cutoff has been added to
$U(\vec\rho)$ in $F_{\rm cs}$ to simulate defects. All of the variables are 
dimensionless and scaled, as described in detail below.

We use overdamped Langevin dynamics so the equation of motion for
$\phi(\vec r, t)$ is
\begin{equation}
\frac{\partial\phi(\vec r, \vec t)}{\partial t} = 
-\frac{\delta F[\phi(\vec r,t)]}{\delta \phi(\vec r,t)} + \zeta(\vec r,t),
\label{e:langevin}
\end{equation}
where the Gaussian noise $\zeta$ is related to the
dimensionless temperature $\theta$ by the fluctuation-dissipation
relation,
\begin{equation}
\langle \zeta(\vec r,t)\zeta(\vec r',t')\rangle = 
2 \theta \delta^2(r - r')\delta(t-t').
\label{e:cfdt}
\end{equation} 

Several predictions have been made~\cite{klsh} about the critical
droplet when the system is quenched to just above the spinodal
temperature, $\tau_s$, so that $\Delta\tau=\tau-\tau_s << 1$. In $d=2$ the nucleation 
barrier is proportional to the Ginzburg parameter 
$G\approx\xi^2 \phi^2/(\theta \chi)\approx\xi^2 \phi^2/(\theta_c \chi)>>1$, 
where $\chi\approx2/\Delta\tau$ is the susceptibility. Because
$\phi\approx\sqrt{\Delta\tau/4} << 1$, the droplet is difficult to
distinguish from the metastable background. For $a>0$,
$\xi\approx \sqrt{a/(4\Delta\tau)}$ and $\tau_s=0$~\cite{klsh}, so 
$G \approx a\tau/(8\theta_c)$. If $a < 0$, 
$\xi\approx\sqrt{|a|/(2\Delta\tau)}$ and $\tau_s=|a|^2/(8b)$, so 
$G\approx|a|\Delta\tau/(4\theta_c)$. The droplet is predicted to 
be modulated at the largest (real) value of the wavenumber
$k_0$ at which the structure function
$S(k)\approx[\tau\pm|a|k^2/4+bk^4/8+\tilde{U}(k)]^{-1}$ diverges. If
$a>0$, $k_0=0$ and the droplet is homogeneous, with $\phi(\vec r)$
in the droplet either everywhere positive or negative. If
$a<0$,
$k_0=\sqrt{|a|/b}$, and the droplet is modulated with wavelength $w=2
\pi
\sqrt{b/|a|}<<\xi$, with alternating regions of positive and negative
$\phi(\vec r)$. 

One of our main goals is to simulate martensites as realistically as
possible. To this end, we need to relate the dimensionless simulation
parameters in Eq.~(\ref{e:GL}) to empirically accessible parameters.
By extending the potential ~\cite{kksw} to include a second
gradient term, the three-dimensional (elastic) free energy near the
critical temperature in terms of measurable quantities is
\begin{eqnarray}
\label{e:Fexp} 
F[\epsilon_3]& = & \!\int_{0}^{L}\! d^3r \left[ \frac{C_1}{2} \epsilon_1^2 
+ \frac{C_2}{2} \epsilon_2^2 + \frac{C_3}{2} \epsilon_3^2 
- \frac{C_4}{4} \epsilon_3^4 \right. \nonumber \\ 
& & \mbox{} \left. + \frac{C_6}{6} \epsilon_3^6 
+ \frac{\kappa_1}{2 a_0^2} (\vec{\nabla}\epsilon_3)^2 
+ \frac{\kappa_2}{4 a_0^4} (\nabla^2 \epsilon_3)^2 \right], 
\end{eqnarray}
where $a_0$ is the crystal lattice spacing, $L$ is the linear dimension of
the system, and $C_1$ through $C_6$, $\kappa_1/a_0^2$, and
$\kappa_2/a_0^4$ are elastic constants in units of $\rm{N/m^2}$. $L/a_0>>\xi$ 
to minimize finite size effects. All the parameters in Eq.~(\ref{e:Fexp}) can 
be determined empirically~\cite{kksw}; for example, 
$\kappa_1/a_0^2$ is determined from the curvature at small $k$ of the phonon 
dispersion curve obtained from neutron scattering experiments~\cite{sato}. We scale all 
the elastic constants by $A_0=9 C_4^3/(128 C_6^2)$ and the strains by 
$\epsilon_0=\sqrt{3 C_4/(4 C_6)}$ to make the two martensite 
minima of the homogeneous part of $F$ in Eq.~(\ref{1b}) near unity,
and we scale all distances by $a_0$. We define the dimensionless
variables 
$\vec{\tilde r}=\vec r/a_0$, 
$\tilde L=L/a_0$, $\tilde F=F/(A_0 a_0^3)$, 
$e_{1,2}=\epsilon_{1,2}/\epsilon_0$, $\phi=\epsilon_3/\epsilon_0$, 
$A_{1,2}=C_{1,2} \epsilon_0^2/A_0$, $\tau = C_3 \epsilon_0^2/(2 A_0)$, 
$a = 2 \kappa_1\epsilon_0^2/(A_0 a_0^2)$, and 
$b=2 \kappa_2 \epsilon_0^2/(A_0 a_0^4)$.

To find the proper scaling for the time, noise, and the temperature, we
use the Langevin equation for $\epsilon_3(\vec r,t)$, which is 
Eq.~(\ref{e:langevin}) with $\phi\rightarrow\epsilon_3$, $F[\phi]\rightarrow
F[\epsilon_3]$, and an explicit friction coefficient $\gamma$
multiplying the time derivative. The dimensionless time and noise
are $\tilde t = A_0 a_0 t/(L \gamma \epsilon_0^2)$ and 
$\tilde \zeta = L \epsilon_0 \zeta/(A_0 a_0)$. The factor of $L$ is a result 
of going from $\epsilon_3(x,y,z,t)$ to $\phi(x,y,t)$. The 
fluctuation-dissipation relation Eq.~(\ref{e:cfdt}) requires that we define
$\theta=k_{\rm{B}} T/(A_0 a_0^3)$. $\theta_c$ and $\theta_0$ are defined
similarly with $T$ replaced by $T_c$, the critical  temperature, and $T_0$ is
the temperature at which the three minima of the homogeneous part of $F$ in
Eq.~(\ref{e:Fexp}) are degenerate. Note that we can change the
effective temperature of the simulations either by changing $T$ or $a_0$. We
drop all tildes in the following.

To discretize the Langevin equation, we used a simple forward Euler 
method~\cite{Press:1992} for the time derivative. Higher order
algorithms take more time and give similar results. The random noise
$\zeta$ is computed by multiplying the standard deviation of the noise
$\sigma_{\zeta}$ by a random number chosen from a Gaussian
distribution with unit variance,
$G_{i,j,\alpha}$~\cite{Press:1992}. On the
lattice
$\delta^2(\vec r - \vec r') \approx 1/\delta x^2$ and 
$\delta(t - t') \approx 1/\delta t$, so $\zeta(\vec
r,t)\approx\sigma_{\zeta}G_{i,j,\alpha}$, where
$\sigma_{\zeta}=\sqrt{2 \theta/(\delta x^2 \delta t)}$. The treatment
of the spatial derivatives and the nonlocal term is not as simple,
however. 

The nonlocal term in the Langevin equation that arises from Eq.~(1e) is
a convolution with the kernel 
$\hat K(\vec k) = \!\int\! d^2\rho\, U(\vec{\rho})e^{-|\vec{\rho}|/R}
e^{-i\vec{k}\cdot \vec{\rho}}$. Here $\hat K(\vec k)$ needs to be computed
only once. The exponential factor $e^{-|\vec{\rho}|/R}$ is computed with 
the origin chosen at the center of the lattice so that it obeys periodic 
boundary conditions. On the discrete lattice the indices $i$ and $j$
correspond to $x$ and $y$, $l$ and $m$ correspond to $k_x$ and 
$k_y$, and $\alpha$ corresponds to $t$. We write
$\phi(x,y,t)\rightarrow\phi_{i,j,\alpha}$, 
$\hat U(k_x,k_y)\rightarrow\hat U_{l,m}$,
$e^{-|\vec{\rho}|/R}\rightarrow{\rm E}_{i,j}$, and 
$\hat K(k_x,k_y)\rightarrow\hat K_{l,m} \approx\ft[{\rm
E}_{i,j}\ft^{-1}(\hat{U}_{l,m})]$, where $\ft$
and $\ft^{-1}$ represent fast Fourier transforms. The
convolution integral is computed as:
\begin{eqnarray}
\!\int_{0}^{L}\!\!\!d^2r'U(\vec \rho)
e^{-|\vec \rho|/R} \phi(\vec r',t)\approx
\ft ^{-1}[ \ft(\phi_{i,j,\alpha}) \hat K_{l,m}].
\end{eqnarray}

The accuracy of our algorithm depends on treating both spatial
derivatives to fourth order in $\delta x$, taking care to
include cross terms in 
$\nabla^4\phi$. If we define 
$\oplus^{(1)}_{i,j} =\phi_{i+1,j}+\phi_{i-1,j}+\phi_{i,j+1}+
\phi_{i,j-1}$,
$\oplus^{(2)}_{i,j} =\phi_{i+2,j}+\phi_{i,j+2}+\phi_{i-2,j} +
\phi_{i,j-2}$, and 
$\otimes_{i,j}=\phi_{i+1,j+1}+\phi_{i+1,j-1}+\phi_{i-1,j+1}+\phi_{i-1,j-1}$, 
then 
\begin{subequations}
\begin{eqnarray}
\nabla^2 \phi_{i,j}& \approx & \frac{1}{12 (\delta
x)^2}[-\oplus^{(2)}_{i,j} +16\oplus^{(1)}_{i,j} -60\phi_{i,j}]
\label{14a}
\\
\nabla^4 \phi_{i,j}& \approx & \frac{1}{(\delta
x)^4}[\oplus^{(2)}_{i,j} -8\oplus^{(1)}_{i,j} +2\otimes_{i,j}+ 20
\phi_{i,j}].\label{14b}
\end{eqnarray}
\end{subequations}
 
We use the values $T_c=268\,\rm{K}$, $T_0 = 290\,\rm{K}$, and $A_2 = 2
A_1$ for all our simulations; these values correspond to
FePd~\cite{kksw} so that our simulations are as realistic as possible. 
We also used $L=64$ and $\delta x=0.5$. Our procedure is to fix $\delta x$ and 
$\delta t=0.01$ and choose the values of the gradient parameters $a$
and $b$ so that the numerical solution is stable. Our numerical
solutions were checked for accuracy by comparing the simulation results
to the exact analytical solution for the linear case without the noise
and nonlocal terms. Numerical stability for the complete equation of
motion was checked by varying
$\delta t$ and
$\delta x$. We limit ourselves to $a$ and $b$ such that the Ginzburg
parameter $G\approx5$ and
$\xi\approx 16=L/4$. We choose $A_1<1$ so that the core of the droplet 
will be more visible. We varied the nucleation rate primarily by changing 
$a_0$.

For $a > 0$ we take $a=6.32$, $b=0.01$, $R=6.4$, 
$\tau=6.17\times 10^{-3}$, $A_1=1\times 10^{-3}$, and
$a_0=2.1544\times 10^{-8}$\,m. FePd has a crystal lattice
spacing of $\approx 3$\,\AA, so this value of $a_0$
corresponds to a coarse-graining factor of 70 and sample size of about
$1.38\,\rm{\mu}$m. The value of $\tau$ corresponds to $T=268.14\,\rm{K}$,
and $a$ corresponds to $\kappa_1/a_0^2=2.79 \times 10^9\,{\rm N/m}^2$. 
The value of $a$ is reasonable compared to the value of $2.5 \times 10^{10}
{\rm N/m}^2$ for FePd quoted in Ref.~\cite{kksw}.

For $a < 0$ we take $a=-1.61$, $b=0.652$, 
$R=4$, $\tau=0.5$, $A_1=0.6$, and $a_0=2.7 \times 10^{-8}\,\rm{m}$. 
These values correspond to the modulation wavelength $w\approx4$, 
$\tau_s=0.49695$, $T=279\,\rm{K}$, and $\Delta \tau=3.048\times 10^{-3}$. 

\begin{figure}[h]
\includegraphics[width=2.7in,angle=0]{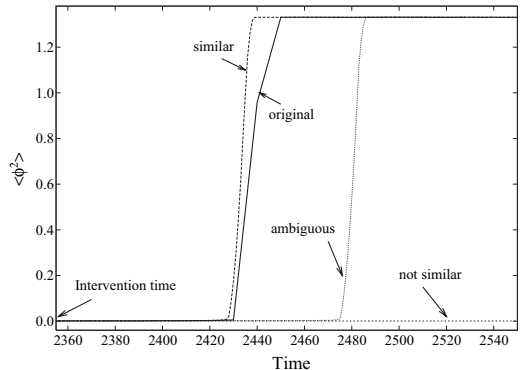}
\caption{Evolution of $\lb \phi^2(t) \rb$ for $a>0$; the original
run and three interventions at $t=2355$. $\lb \phi^2(t) \rb$ of the
original run grows rapidly from its metastable value to the stable
value at 
$t\approx2430$. If the intervention run is similar to the original run,
the same droplet is assumed to have grown; otherwise, it is assumed 
that the same droplet did not grow. Some interventions are ambiguous. 
In this run the intervention time of 2355 is close to the
estimated nucleation time of 
$t_{\rm n}^{+}=2356\pm 35$.}
\label{fig:trialspsb}
\end{figure}

We begin the simulation with $\phi(\vec r)$ chosen at random around the
metastable austenite minimum at $\phi=0$. Within a short time,
$\phi(\vec r)$ equilibrates at the chosen temperature. On a 
time scale much longer than the initial equilibration, the thermal 
noise causes a critical droplet to appear. By looking at the
evolution of the spatial average, $\lb \phi^2(t)\rb$, we can make an
estimate of the nucleation time. From Fig.~\ref{fig:trialspsb} 
we see that the system went from a metastable state with 
$\lb \phi^2 \rb \sim 0$ to the stable state with $\lb \phi^2 \rb \sim 1$ 
at the time $t=2430$. Because we expect the amplitude of the droplet to be 
close to that of the metastable phase, we expect the nucleation time to be 
before the rapid increase in $\lb \phi^2 \rb$.

To determine the nucleation time more precisely, we use an intervention
technique~\cite{mk}. Because the nucleation droplet is a saddle
point~\cite{uk,mk}, the droplet has an equal probability of growing
to the stable state or shrinking back to the metastable state if we 
perturb the system at the nucleation time. We can use this saddle
point property to help us find the nucleation time. We restart the
simulation at the estimated nucleation time, and integrate the 
equations of motion using a new sequence of random numbers for the
thermal noise. Our criterion is that if $8 \pm 4$ of the 16 runs
with different random number sequences show that $\lb \phi^2 \rb$ 
grows at roughly the same time as in the original run, then the
intervention time is equal to the nucleation time. We can then look
at snapshots of $\phi(x,y)$ to see if we can identify the droplet.

For $a>0$, the nucleation time is $t_{\rm n}^{+}=2356 \pm 35$,
corresponding to Fig.~\ref{fig:trialspsb}. A snapshot of 
$\phi(x,y)$ at $t=2356$ is shown in Fig.~2a. Even though
only the core of the droplet is visible above the noise, we can see
that the droplet is homogeneous, as predicted.

\begin{figure*}
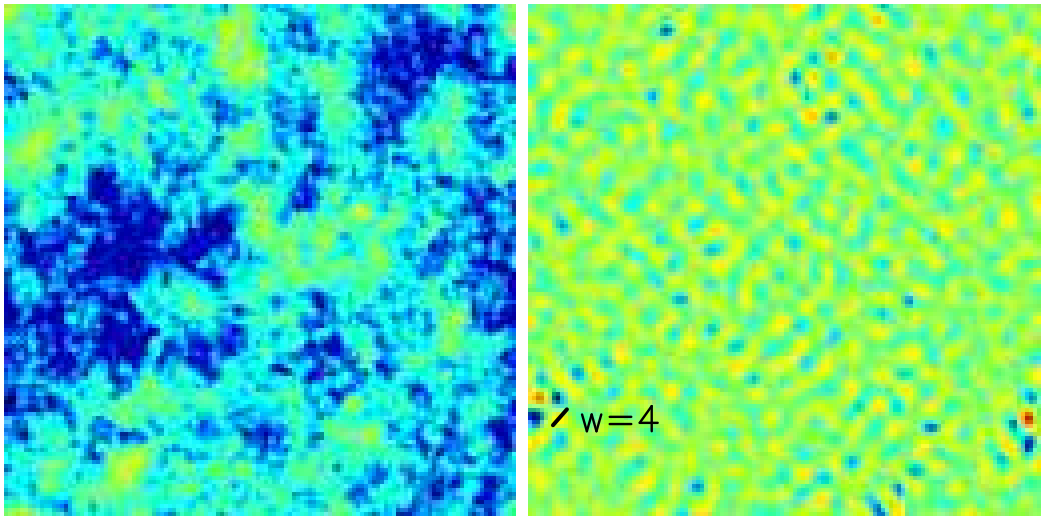

\includegraphics[height=2.7in]{aposplain}\label{fig:2a}
\includegraphics[height=2.7in]{anegphiann}\label{fig:2b}
\caption{(Color) From left:  (a)  The deviatoric strain $\phi(x,y)$ for 
$a>0$ at the estimated nucleation time $t_{\rm n}^{+}=2356$. 
The amplitude of $\phi$ ranges from $-0.08$ (dark blue) to $0.025$ (yellow). 
The part of the droplet visible above the noise is clearly homogeneous, 
showing no modulations between the two low temperature minima (red and blue).
(b)~ The deviatoric strain $\phi(x,y)$ for $a<0$ at the estimated nucleation time $t_{\rm n}^{-}=1236.5$. The amplitude of 
$\phi$ ranges from 
$-0.2$ (dark blue) to 0.2 (red). In the part of the droplet
that is visible, the modulations between the low temperature minima 
(red and blue) have the wavelength predicted by Ref.~\cite{klsh} to 
be $w\approx4$.}
\end{figure*}

For $a<0$, the nucleation time is estimated to be
$t_{\rm n}^{-}=1236.5\pm 1.5$. The error in the nucleation time is 
much less than for $a>0$ because the slower growth in $\lb \phi^2 \rb$
for $a<0$ makes fewer of the interventions ambiguous. A snapshot of the
field at $t=1236.5$ is shown in Fig.~2b. As predicted
in Ref.~\cite{klsh}, the droplet has modulations with 
wavelength $w \approx 4$. This modulation is different from 
the twinning in the stable phase which occurs at wavelength 
$\lambda \sim \sqrt{L}=8$~\cite{slsb}.

These results are important for nucleation in many functional 
materials with long-range interaction in which strain couples to 
some other physical variable, e.g. polarization (in ferroelectrics), 
magnetization (in magnetoelastics), and other multiferroics.  We 
found that in elastic systems the description of nucleation is 
subtle due to the presence of bulk/interface elastic compatibility 
contraints that are manifested as long-range interactions. 
In summary, when the austenite is quenched to near the pseudospinodal,
the structure of the nucleation droplet is different than the
structure of the stable martensite phase. If the curvature of the
phonon dispersion curve at small $k$ is positive, then $a>0$ and the
droplet is homogeneous. If the curvature is negative, then $a<0$ 
and the droplet is modulated with a wavelength $w$. The parameters 
we used are consistent with austenite to martensite transitions 
in FePd. We conclude, therefore that the classical nucleation
picture is not applicable to these transitions and that the spinodal 
nucleation scenario is the better approach to understanding these 
transitions.
 
\section*{Acknowledgments}

C.J.G.\ would like to thank Daniel Blair for his assistance with Fig. 2, and LANL where some of this work 
was done. W.\ K.\ acknowledges support from ASC Materials Modelling Program
at LANL. Work at LANL was supported by the U.S. Department of Energy.

\end{document}